%% file: draft_v17.tex
\documentclass[prd,amsmath,amssymb,showpacs,superscriptaddress,nofootinbib,twocolumn]{revtex4-1}
\usepackage{graphicx}
\usepackage{epsfig,graphics,subfigure,psfrag,amsmath,amssymb}
\usepackage{lineno}
\usepackage{dcolumn}
\usepackage{bm}
\usepackage{overpic}
\usepackage{xspace}
\usepackage{xcolor}
\usepackage{rotating}
\usepackage{color}
\usepackage{multirow}
\usepackage{colortbl}
\usepackage{epstopdf}
\usepackage{float}
\usepackage{makecell}
\usepackage[colorlinks,linkcolor=blue,anchorcolor=blue,citecolor=blue]{hyperref}
\newcommand{\BR}{{\cal B}}

\newcommand{\jpsi}{J/\psi}

\newcommand{\BESIIIorcid}[1]{\href{https://orcid.org/#1}{\hspace*{0.1em}\raisebox{-0.45ex}{\includegraphics[width=1em]{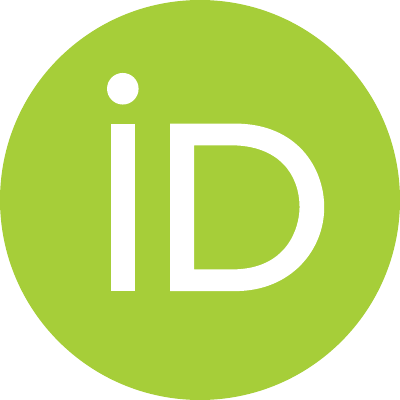}}}}

\newcommand{\BESIII}{BES\uppercase\expandafter{\romannumeral3}\xspace}

\begin{document}
\title{\boldmath Amplitude analysis and branching fraction measurement of $\jpsi\to \Lambda\bar{\Sigma}^0\eta +\mathrm{c.c}$}
\author{
\begin{small}
  \begin{center}
    \input{authorlist_2025-09-11}
\end{center}
\end{small}
}

\date{\today}

\begin{abstract}
Based on a sample of $(10087\pm44)\times10^{6}$ $J/\psi$ events collected with the BESIII detector, a partial-wave analysis of $ \jpsi \to \Lambda \bar{ \Sigma }^0\eta +\mathrm{c.c} $ is performed for the first time. 
The dominant contributions are found to be excited $\Lambda$ states with $J^P=1/2^-$ and  $J^P=1/2^+$ in the $\eta\Lambda$ mass spectra.
The measured masses and widths are $M=1668.8\pm3.1\pm21.2$ MeV/$c^2$ and $\Gamma=52.7\pm4.2\pm17.8$ MeV for the $\Lambda(1670)$, and $M=1881.5\pm16.5\pm20.3$ MeV/$c^2$ and $\Gamma=82.4\pm18.2\pm8.9$ MeV for the $\Lambda(1810)$, respectively.
The branching fraction is determined to be $ \mathcal{B}(\jpsi \to \Lambda \bar{ \Sigma }^0\eta +\mathrm{c.c}) $ = $(3.44 \pm 0.11 \pm 0.13) \times 10^{-5}$. The first uncertainties are statistical and the second systematic. 
\end{abstract}

\maketitle

\section{INTRODUCTION}

The spectra and structure of excited light-flavor baryons provide information that can improve our understanding of confinement, chiral symmetry breaking, and other nonperturbative aspects of Quantum Chromodynamics~\cite{Kamano:2015hxa}.
Due to their unstable nature, excited baryons strongly couple to meson-baryon continuum states, thereby forming resonant structures such as the nucleon resonances ($N^*$, $\Delta^*$) with strangeness $S$=0 and excited hyperon resonances ($\Lambda^*$, $\Sigma^*$) with strangeness $S$=1.  These systems therefore serve as key probes into the underlying QCD dynamics~\cite{Kamano:2015hxa}.

While the quark model predicts numerous $\Lambda^*$ baryon resonance states,
the $\Lambda(1670)$, which has a mass near the $\eta\Lambda$ threshold, has inspired a wide range of theoretical interpretations, including: 
the SU(3) octet partner of the $N(1535)$~\cite{Koniuk:1979vy}, a $K\Xi$ bound state using a meson-baryon framework~\cite{Oset:2001cn}, and a mixture of three-quark and five-quark components~\cite{Zou:2013af}. 
However, experimental information remains sparse. 
The interpretation of $\bar{K}N$ scattering data relies on model-dependent partial-wave analysis 
(PWA)~\cite{Kamano:2015hxa,Zhang:2013sva}, and the Belle Collaboration observed $\Lambda(1670)$ in the 
decay $\Lambda_c^+\to\eta\Lambda\pi^+$ based on only a one-dimensional fit~\cite{Belle:2020xku}. 
Regarding the $\Lambda(1810)$ resonance, there are currently different viewpoints on its existence. One point of view is that its existence is very likely but requires confirmation, with its quantum numbers and branching fractions poorly determined in the Particle Data Group (PDG)~\cite{ParticleDataGroup:2024cfk}. Another point of view is that the quark model provides only weak evidence for its existence~\cite{Sarantsev:2019xxm}.
Observations of the $\Lambda(1810)$ in the mass distributions of $N\bar{K}$, $\Sigma\pi$, and $\Lambda\sigma$ and searches for the $\Lambda(1810)$ in the mass distribution of $\eta\Lambda$ is thus crucial for clarifying its structural properties.

The charmonium vector meson $J/\psi$ is an SU(3) singlet $c\bar{c}$ bound state with isospin $I = 0$.
It provides a unique laboratory for baryon spectroscopy and investigations of SU(3) flavor symmetry. 
The $\jpsi \to \Lambda \bar{\Sigma}^0 \eta $ decay is an isospin-violating process, which offers a good opportunity for the study of baryonic states. 
At the level of $J/\psi$ decays, isospin violation can usually be ascribed to electromagnetic decays, such as in the cases of $J/\psi\rightarrow\rho\eta$, $\jpsi\to\bar{\Lambda}\Sigma^0$, and $\jpsi\to\bar{\Lambda}(1520)\Sigma^0$~\cite{BESIII:2012xdg,DM2:1988bfq}. In the decays of baryons, isospin violation could happen in either weak or electromagnetic decays, such as in $\Lambda\rightarrow p\pi$ and $\Sigma\rightarrow\gamma p$.  However, the dominant contribution to baryon the excited baryon decays comes from the strong interactions, where isospin is conserved. 

The high production rate of baryons in charmonium decays presents an excellent opportunity for the study of excited baryons. By analyzing a sample of $(10087\pm 44) \times 10^6$ $\jpsi $ events collected with the BESIII detector~\cite{BESIII:2016kpv}, we perform a PWA of the decay $\jpsi \to \Lambda \bar{\Sigma}^0 \eta $ to measure its branching fraction and to investigate excited states of the $\Lambda$.  Charge-conjugate processes are implied throughout.

\section{BESIII DETECTOR AND MONTE CARLO
 SIMULATION}
The BESIII detector~\cite{BESIII_2009_detector} records symmetric $e^+e^-$ collisions 
provided by the BEPCII storage ring~\cite{Yu:IPAC2016-TUYA01}
in the center-of-mass energy range from 1.84 to 4.95~GeV,
with a peak luminosity of $1.1 \times 10^{33}\;\text{cm}^{-2}\text{s}^{-1}$ 
achieved at $\sqrt{s} = 3.773\;\text{GeV}$. 
BESIII has collected large data samples in this energy region~\cite{BESIII:2020nme}. The cylindrical core of the BESIII detector covers 93\% of the full solid angle and consists of a helium-based multilayer drift chamber~(MDC), a plastic scintillator time-of-flight
system~(TOF), and a CsI(Tl) electromagnetic calorimeter~(EMC),
which are all enclosed in a superconducting solenoidal magnet
providing a 1.0~T magnetic field.
The magnetic field was 0.9~T in 2012, which affects 10\% of the total $J/\psi$ data.
The solenoid is supported by an
octagonal flux-return yoke with resistive plate counter muon
identification modules interleaved with steel. 
The charged-particle momentum resolution at 1 $\mathrm{GeV}/c$ is
$0.5\%$, and the ${\rm d}E/{\rm d}x$
resolution is $6\%$ for electrons
from Bhabha scattering. The EMC measures photon energies with a
resolution of $2.5\%$ ($5\%$) at $1$~GeV in the barrel (end cap)
region. The time resolution in the TOF barrel region is 68~ps, while
that in the end cap region was 110~ps. 
The end cap TOF system was upgraded in 2015 using multigap resistive plate chamber technology, providing a time resolution of
60~ps~\cite{BESIII_2019_detector}, which benefits 87$\%$ of the data used in this analysis.

Simulated data samples produced with a GEANT4-based~\cite{GEANT4:2002zbu} Monte Carlo (MC) package, which includes the geometric description of the BESIII detector and the detector response, are used to determine detection efficiencies and to estimate backgrounds. The simulation
 models the beam energy spread and initial state radiation in the $e^+e^-$ annihilations with the generator KKMC~\cite{Jadach:2000ir,Jadach:1999vf}. All particle decays are modelled with EVTGEN~\cite{Lange:2001uf,Ping:2008zz} using branching fractions either taken from the PDG~\cite{ParticleDataGroup:2024cfk}, when available, or
 otherwise estimated with LUNDCHARM~\cite{Chen:2000tv,Yang:2014vra}.  Final state radiation from charged final state particles is incorporated using the PHOTOS package~\cite{Richter-Was:1992hxq}. 
 Possible backgrounds are studied using a sample of $J/\psi$ inclusive events in which the known decays of the $J/\psi$ are modeled with branching fractions set to be the world average values from the PDG~\cite{ParticleDataGroup:2024cfk}. Signal MC samples are generated uniformly in phase space.

\section{Event selection}

The decay $\jpsi \to \Lambda \bar{\Sigma}^0 \eta$ 
is studied using the subsequent decays $\Lambda\to p\pi^-$, $\bar{\Sigma}^0\to\gamma\bar{\Lambda}$, $\bar{\Lambda}\to\bar{p}\pi^+$, and  $\eta\to\gamma\gamma$, resulting in the final state $\gamma\gamma\gamma p\bar{p} \pi^+\pi^-$. Each event is thus required to contain at least three good photon candidates and four charged track candidates with net charge zero. Charged tracks detected in the MDC are required to be within a polar angle $(\theta)$ range of $|\cos \theta| \leq 0.93$, where $\theta$ is defined with respect to the $z$-axis, which is the symmetry axis of the MDC.
 
Photons are reconstructed from showers in the EMC with a deposited energy of at least 50 MeV in the barrel region ($|\cos \theta| < 0.8$) and 50 MeV in the end cap regions $(0.86 < |\cos \theta|< 0.92)$. The opening angle between the shower position and the charged tracks extrapolated to the EMC must be greater than 10 degrees. Finally, photons are required to arrive within 700 ns from the event start time to reduce background from photons that do not originate from the same event. 

The $\Lambda$ and $\bar{\Lambda}$ candidates are reconstructed by combining pairs of oppositely charged tracks with pion and proton mass hypotheses and by imposing a secondary vertex constraint~\cite{Xu:2009zzg,BESIII:2021cvv}. Events with at least one $p\pi^-$ candidate for the $\Lambda$ decay and one $\bar{p}\pi^+$ candidate for the $\bar{\Lambda}$ are retained. 
In the case of multiple $\Lambda\bar{\Lambda}$ pair candidates, the one with the minimum value of $\chi_{\mathrm{svtx}}^2(\Lambda) + \chi_{\mathrm{svtx}}^2(\bar{\Lambda})$ is chosen, where $\chi_{\mathrm{svtx}}^2(\Lambda)$ and  $\chi_{\mathrm{svtx}}^2(\bar{\Lambda})$ are the $\chi^2$ of the secondary vertex  fits for the $\Lambda$ and $\bar{\Lambda}$, respectively. 
To improve the momentum and energy resolution and to reduce background contributions, a four-constraint (4C) energy-momentum conservation kinematic fit is applied to the event candidates under the hypothesis of $\Lambda\bar{\Lambda}\gamma\gamma\gamma$. For events with more than three photons, the combination with the best fit quality is selected. To reject possible background events from $J/\psi\rightarrow \Lambda\bar{\Lambda}\gamma\gamma$ and  $J/\psi\rightarrow \Lambda\bar{\Lambda}\gamma\gamma\gamma\gamma$, we require the $\chi^2$ of fits to the $\Lambda\bar{\Lambda}\gamma\gamma$ and  $\Lambda\bar{\Lambda}\gamma\gamma\gamma\gamma$ hypotheses be greater than the $\chi^2$ for the $\Lambda\bar{\Lambda}\gamma\gamma\gamma$ hypothesis.

After satisfying the above requirements, a five-constraint (5C) kinematic fit is performed, incorporating energy-momentum conservation and a mass constraint on the $\eta\to \gamma\gamma$ decay~\cite{ParticleDataGroup:2024cfk}. 
The combination with the smallest $\chi^2_{\rm{5C}}(\Lambda\bar{\Lambda}\eta(\gamma\gamma)\gamma)$ is kept for further analysis, and we require $\chi^{2}_{\rm{5C}}(\Lambda\bar{\Lambda}\eta(\gamma\gamma)\gamma)<50$ for the 
$J/\psi\to\Lambda\bar{\Sigma}^0\eta$ candidates.

To suppress background events from $ \jpsi \to \Lambda \bar{\Lambda}\pi^0\gamma$, another 5C kinematic fit is performed under the hypothesis of $\Lambda\bar{\Lambda}\pi^0\gamma$, incorporating a constraint on the four-momentum of the final state to that of the initial $J/\psi$ (4C) and an additional mass constraint (1C) on the $\pi^0\to\gamma\gamma$ decay.
The $\chi^2_{\mathrm{5C}}$ of the 5C kinematic fit under the hypothesis of $\Lambda\bar{\Lambda}\eta\gamma$ is required to be less than that of $\Lambda \bar{\Lambda} \pi^0 \gamma$.

\begin{figure}[htbp]
    \centering
    \begin{overpic}[width=0.38\textwidth]{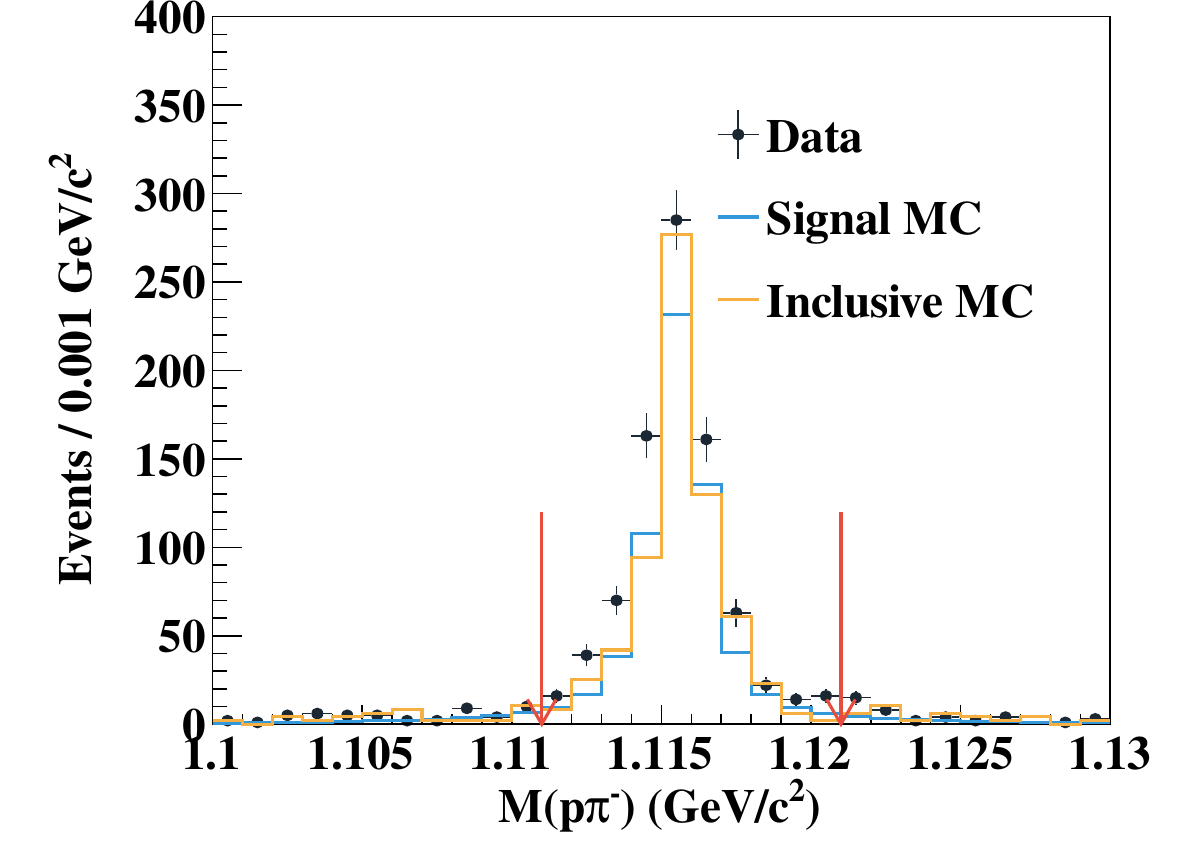}
        \put(22,57){$(a)$}
    \end{overpic}
    \begin{overpic}[width=0.38\textwidth]{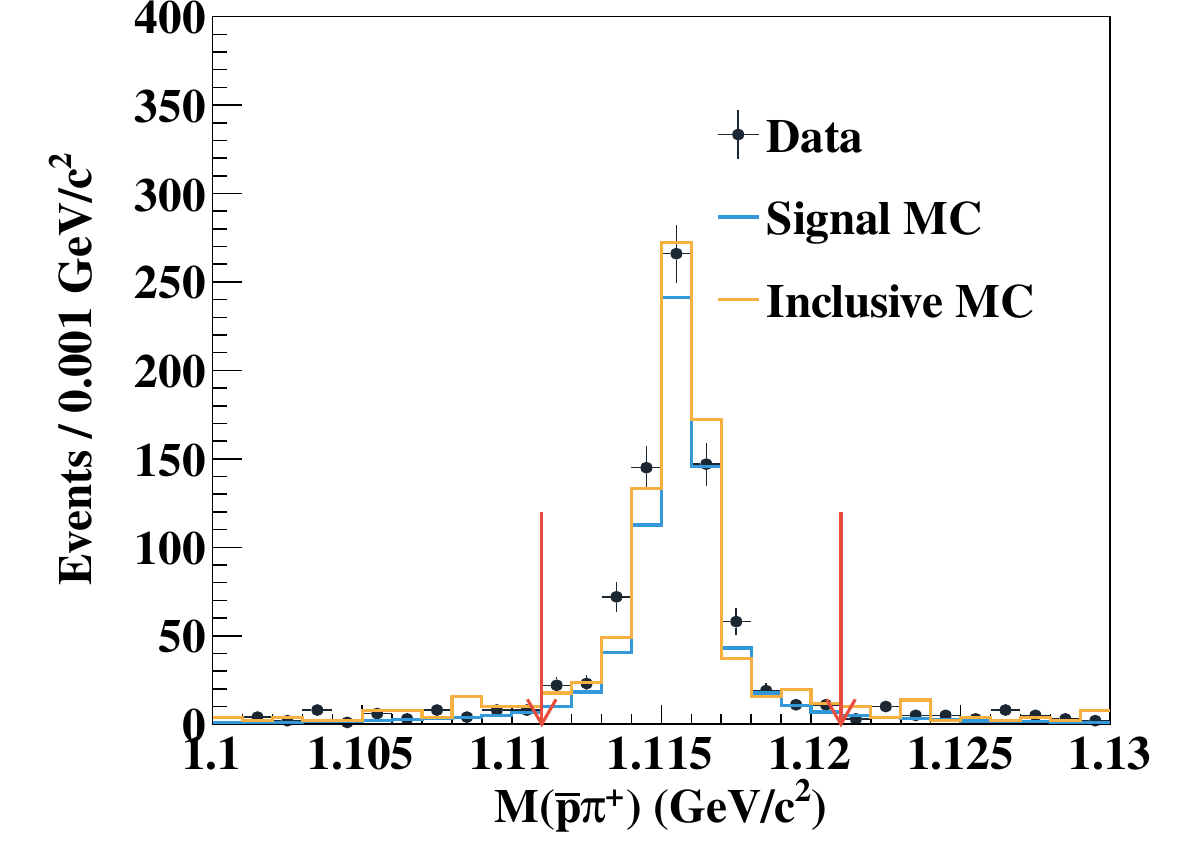}
        \put(22,57){$(b)$}
    \end{overpic}
    \caption{The distributions of (a) $ M ( p \pi^- $) and (b) $ M ( \bar{p} \pi ^+)$. Dots with error bars represent data, the blue line is the simulated signal shape, and the yellow line denotes the background from the inclusive MC sample. The red arrows denote the $\Lambda$ and $\bar{\Lambda}$ mass windows.}
    \label{line_Lambda}
\end{figure}

After applying the above selection criteria, the invariant mass distributions of $p\pi^-$ and $\bar{p}\pi^+$
are shown in Fig.~\ref{line_Lambda}. 
We require $1.111 < M(p\pi^- / \bar{p}\pi^+) < 1.121$ $\mathrm{GeV}/c^2$ to 
select $\Lambda$ and $\bar{\Lambda}$ candidates.
To suppress background arising from miscombinations of the charge-conjugated channel, the invariant mass of $\gamma\Lambda$ from $J/\psi\to\Lambda\bar{\Sigma}^0\eta$ is required to not be in the $\Sigma^0$ signal region $M(\gamma\Lambda) \notin (1.179,2.204)$~$\mathrm{GeV}/c^2$.

\begin{figure}[htbp]
    \centering
    \begin{overpic}[width=0.38\textwidth]{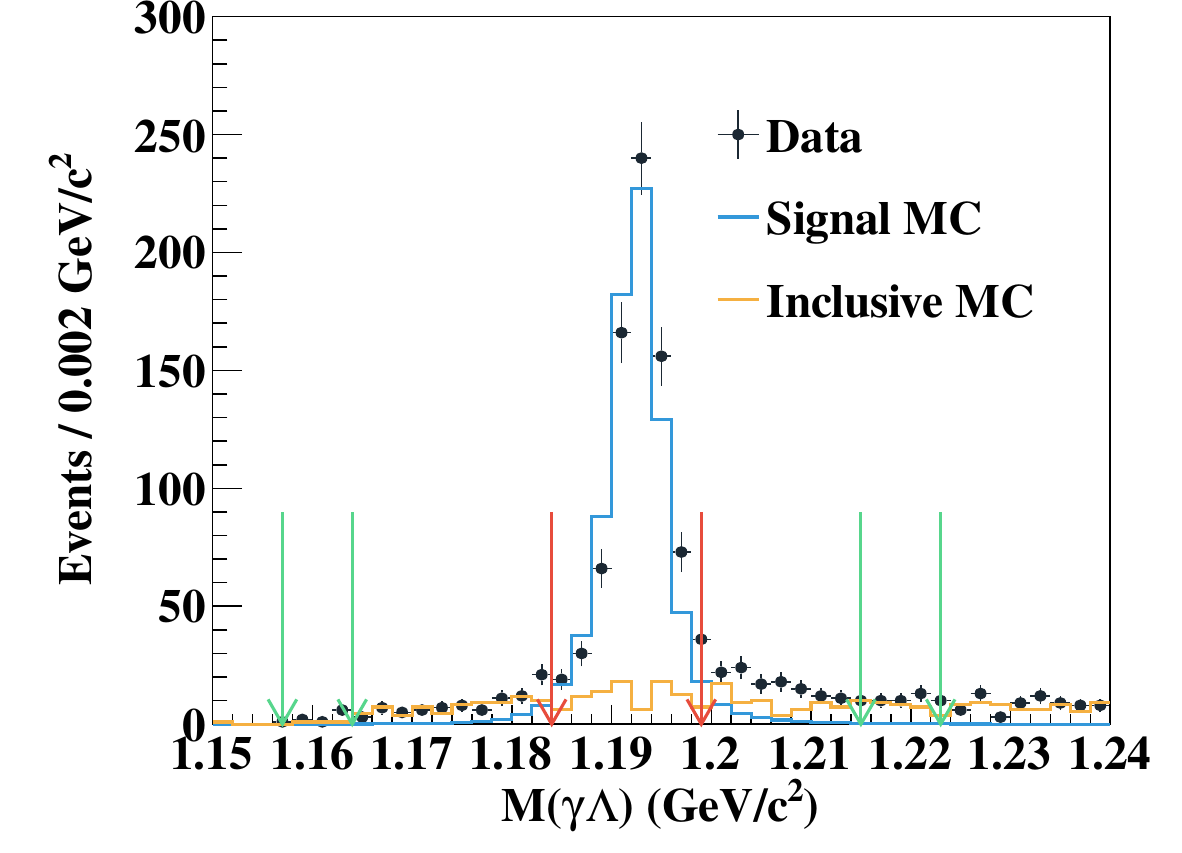}
        \put(22,57){$(a)$}
    \end{overpic}
    \begin{overpic}[width=0.38\textwidth]{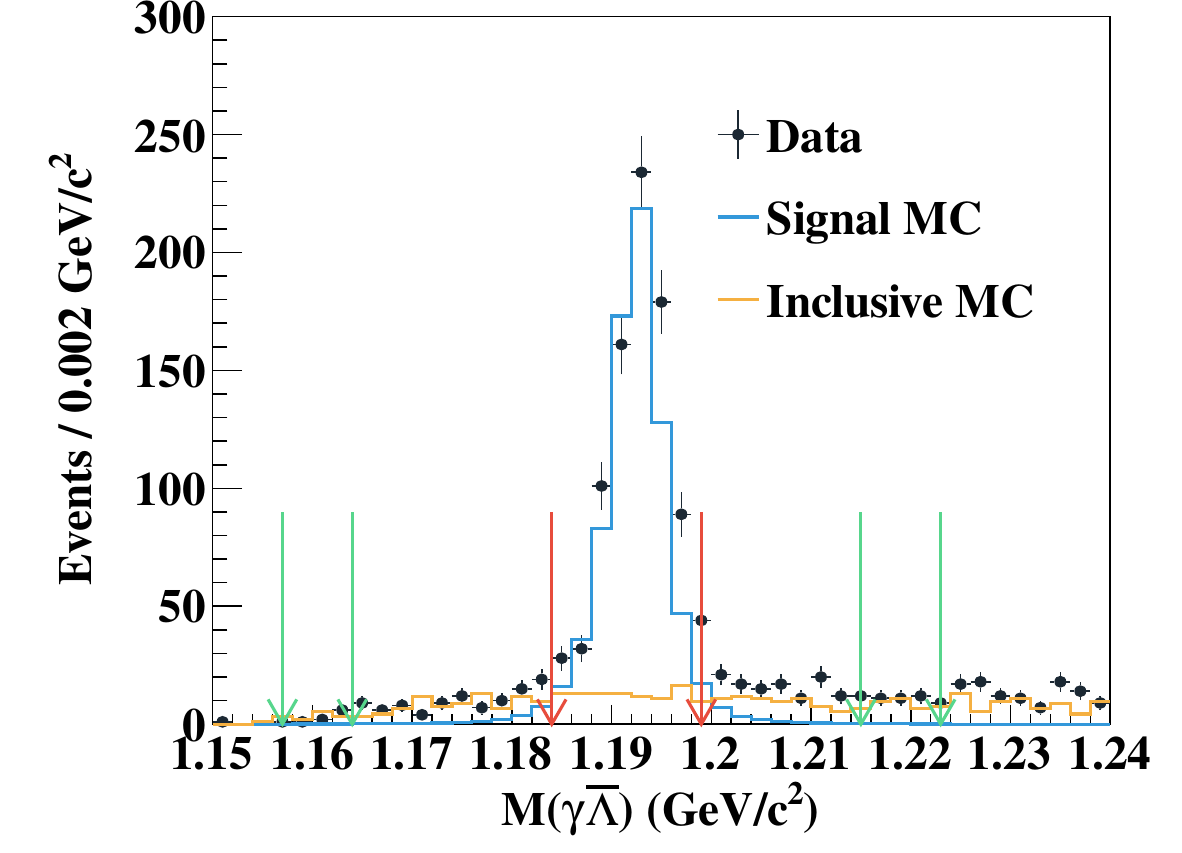}
        \put(22,57){$(b)$}
    \end{overpic}
    \caption{The distributions of (a) $ M (\gamma\Lambda $) and (b) $ M (\gamma\bar{\Lambda})$. Dots with error bars represent data, the blue line is the simulated signal shape, and the yellow line denotes the background from the inclusive MC sample. The $\Sigma^0$ and $\bar{\Sigma}^0$ signal region is shown with the red arrows and $\Sigma^0$ and $\bar{\Sigma}^0$ sideband ranges are shown with the short green arrows.}
    \label{line_sigma}
\end{figure}

To investigate possible background contributions, the same selection criteria are applied to an inclusive MC sample of 10 billion $J/\psi$ events. The background channels exhibit smooth distributions without significant peaking structures around the nominal $\bar{\Sigma}^0$ mass, as shown in Fig.~\ref{line_sigma}.

An off-resonance data sample of $167.06$ $\mathrm{pb}^{-1}$ taken at a center-of-mass energy of $3.08$ $\rm{GeV}$ is used to estimate the possible background from continuum processes~\cite{BESIII:2012pbg}. After applying the same selection criterion as for $J/\psi$ data, only two events survive. Therefore, the background from continuum processes is ignored and possible interference effects are also ignored.  This is expected to have a negligible impact on the branching fraction measurement.

\section{PARTIAL WAVE ANALYSIS}

\begin{figure}[htbp]
    \centering
    \begin{overpic}[width=0.38\textwidth]{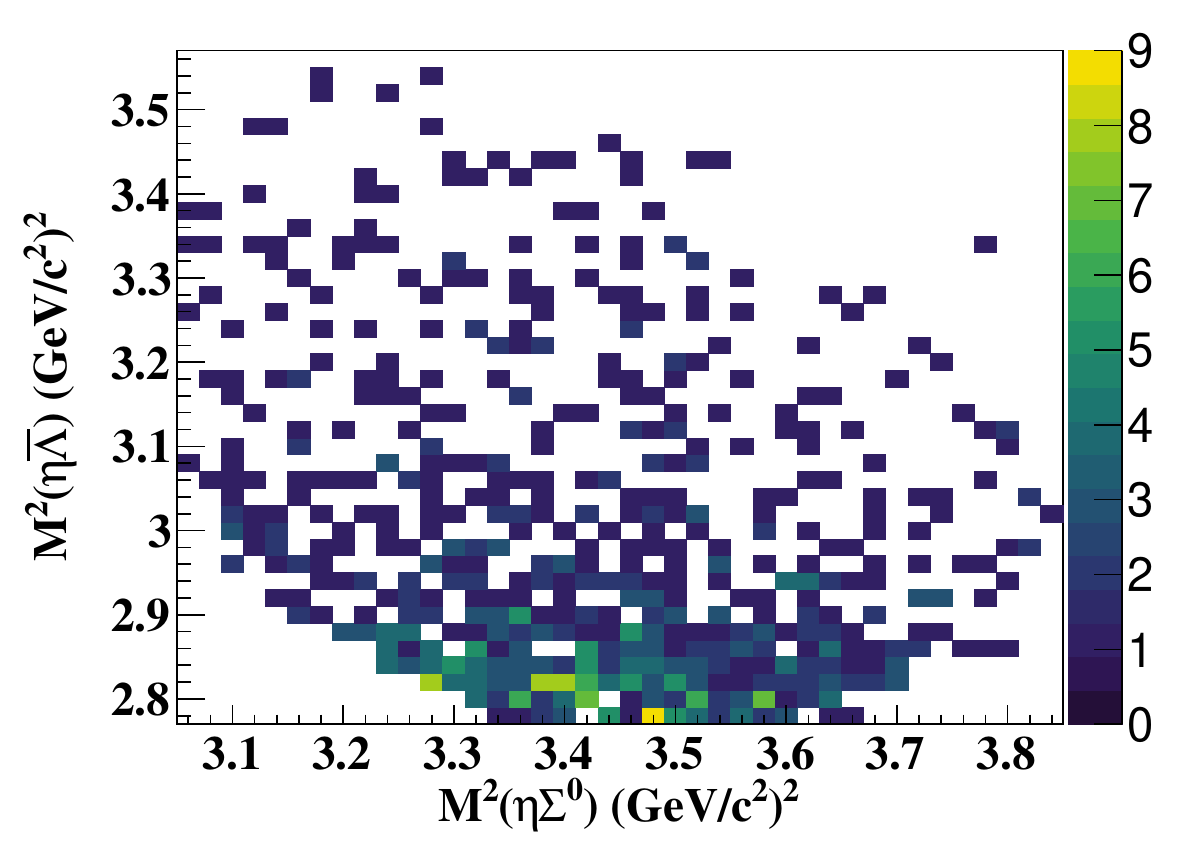}
        \put(70,60){$(a)$}
    \end{overpic}
    \begin{overpic}[width=0.38\textwidth]{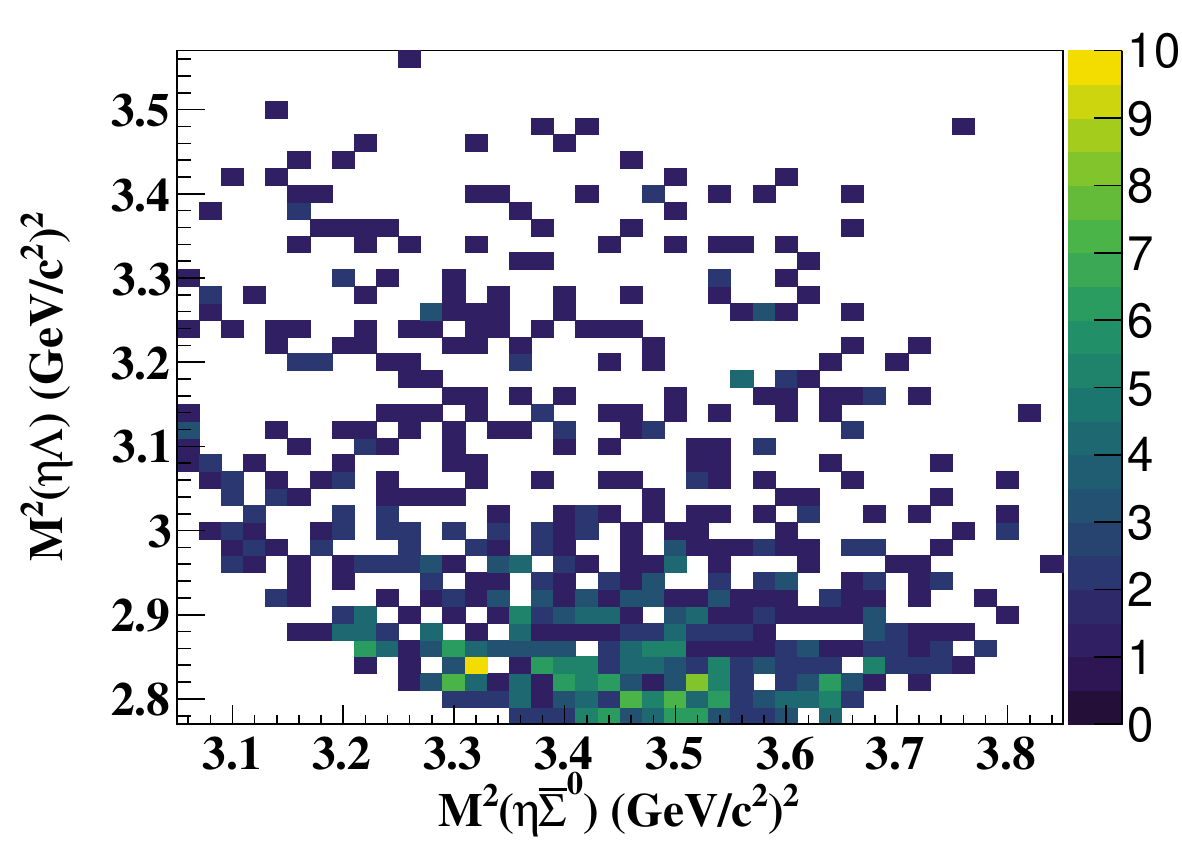}
        \put(70,60){$(b)$}
    \end{overpic}
    \caption{The Dalitz plots of (a) $J/\psi \to\bar{\Lambda}\Sigma^0\eta $ and (b) $J/\psi \to\Lambda\bar{\Sigma}^0\eta $ events in the $\Sigma$ signal  region from data.}
    \label{Dalitz}
\end{figure}

The Dalitz plots in Fig.~\ref{Dalitz} show $M^2(\eta\Lambda) $ versus $M^2(\eta\bar{\Sigma}^0)$ for the decay $\jpsi \to \Lambda \bar{\Sigma}^0 \eta$ and 
$M^2(\eta\bar{\Lambda}) $ versus $M^2(\eta\Sigma^0)$ for the charge-conjugated mode, where the $\Lambda^*$ resonances are visible as horizontal bands. 
An additional requirement on the $\gamma\bar{\Lambda}$ invariant mass, $M(\gamma\bar{\Lambda})\in (1.184, 1.199)$ $\mathrm{GeV}/c^2$, is applied to improve the signal purity, as shown in Fig.~\ref{line_sigma}, where the red arrows are used for the signal definition. A total of 1617 
candidates are retained for the subsequent PWA. The background in the signal region is estimated using events 
in the $\bar{\Sigma}^0$ sideband regions, 
defined as $1.157<M(\gamma\bar{\Lambda}) <1.164$~$\mathrm{GeV}/c^2$ and 
$1.215<M(\gamma\bar{\Lambda}) <1.223$~$\mathrm{GeV}/c^2$, yielding $N_{\Sigma}^{\mathrm{side}}=140\pm12$ events, as shown in Fig.~\ref{line_sigma}, where the green arrows are used for the sidebands definition.
The decay amplitude is constructed using the helicity amplitude formalism, and the full procedure is implemented based on the open-source framework TF-PWA~\cite{Jiang:2024vbw}.

\subsection{Analysis method}
To construct the decay amplitude of $J/\psi \to\Lambda\bar{\Sigma}^0\eta$, the helicity formalism is used based on the isobar model describing the three-body decay as a two-step sequential quasi-two-body decay. For each two-body decay $A\to B+C$, the helicity amplitude can be written as
\begin{equation}
    A^{A \rightarrow B+C}_{\lambda_{A},\lambda_{B},\lambda_{C}} = H_{\lambda_{B},\lambda_{C}}^{A \rightarrow B+C} D^{J_{A}\star}_{\lambda_{A},\lambda_{B}-\lambda_{C}}(\phi,\theta,A).
\end{equation}
In the Wigner D-function, $D^{J_{A^*}}_{\lambda_{A},\lambda_{B}-\lambda_{C}}(\phi,\theta,A)$, the $\phi$ and $\theta$ represent the polar and azimuthal angles, respectively, of the momentum of particle B defined in the rest frame of particle A. 
The definitions can be found in Ref.~\cite{Wang:2020giv}. The amplitude $H_{\lambda_{B},\lambda_{C}}^{A \rightarrow B+C}$ is given by the LS coupling formula~\cite{Chung:1997jn} along with barrier factor terms
\begin{equation}
\begin{aligned}
H_{\lambda_{B},\lambda_{C}}^{A \rightarrow B+C} &=
\sum_{ls} g_{ls} \sqrt{\frac{2l+1}{2 J_{A}+1}} \langle l_0; s \delta|J_{A} \delta\rangle \\
&\langle J_{B} \lambda_{B} ;J_{C} -\lambda_{C} | s \delta \rangle q^{l} B_{l}'(q, q_0, d),
\end{aligned}
\end{equation}
where $g_{ls}$ is the fitting parameter, $J_{A,B,C}$ are the spins of the particles $A$, $B$ and $C$, $\lambda_{B,C}$ are the helicities for the particles $B$ and $C$, and $\delta = \lambda_B - \lambda_C$ is the helicity difference. $B_{l}'(q, q_0, d)$ is the reduced Blatt-Weisskopf barrier factors~\cite{ParticleDataGroup:2024cfk,Chung:1993da,Chung:1995dx}, the radius $d$ is chosen as $d$ = 0.73 fm, which is the same as in Ref.~\cite{BESIII:2019dme}. The normalization factor $q_0$ is calculated at the nominal resonance mass.

The amplitude for a complete decay chain is constructed as the product of each two-body decay amplitude and the resonant propagator $R$. For example, in the sequential decay $A\to R+B$ with $R \to C+D$, the amplitude is written as
\begin{equation}
A^{A \rightarrow R+B,R \rightarrow C+D}_{\lambda_{A},\lambda_{B},\lambda_{C},\lambda_{D}} = \sum_{\lambda_{R}}A^{A \rightarrow R+B}_{\lambda_{A},\lambda_{R},\lambda_{B}}{R(m_{R})}\color{black} A^{R \rightarrow C+D} _{\lambda_{R},\lambda_{C},\lambda_{D}}.
\end{equation}
The propagator $R$ includes different models. The Breit-Wigner formula is taken as
\begin{equation}
    R(m) = \frac{1}{m^2_0 - s -im_{0}\Gamma},
\end{equation}
where $m_0$ and $\Gamma$ are the mass and width of resonances, and $m=\sqrt{s}$ is the mass of $\eta\Lambda$.

The probability density to observe the $i$-th event which is denoted by its kinematic variables $\xi_i$ is
\begin{equation}
    P(\xi_{i})=\frac{\omega(\xi_i)\epsilon(\xi_i)}{\int \omega(\xi) \epsilon(\xi)\, d\xi},
\end{equation}
where $\omega(\xi_i)$ is the differential cross-section, $\epsilon(\xi_i)$ is the detection efficiency,
\begin{equation}
    \omega(\xi)\equiv \frac{d\sigma}{d\Phi}=\left|\sum_{j}A_{j}\right|^2,
\end{equation}
where $A_j$ is the amplitude of the $j$-th process, $d\Phi$ is the standard element of phase space, $\sigma \equiv \int \omega(\xi) \epsilon(\xi)\, d\xi$ is the measured total cross section.

The joint probability for observing the $N$ events in the data sample is:
\begin{equation}
\begin{aligned}
    \mathcal{L} \equiv P(\xi_1,\xi_2\cdots,\xi_n) 
    &= \prod^{N}_{i=1} P(\xi_i) \\
    &= \prod^{N}_{i=1}\frac{\omega(\xi_i)\epsilon(\xi_i)}{\int \omega(\xi) \epsilon(\xi) d\xi}.\\
\end{aligned}
\end{equation}

The detection efficiency function $\epsilon(\xi)$ in the above expression can be separated by taking the negative natural logarithm
\begin{equation}
\begin{aligned}
    -\ln\mathcal{L} &\equiv -\ln P(\xi_1,\xi_2\cdots,\xi_n) \\
    &= -\sum_{i=1}^{N}\ln(\frac{\omega(\xi_i)}{\int \omega(\xi) \epsilon(\xi) d\xi}) - \sum_{i=1}^{N}\ln\epsilon(\xi_i).
\end{aligned}
\end{equation}

For the determined MC and data samples, the detection efficiency $\epsilon(\xi_i)$ can be regarded as a constant and has no effect on the fitting results, so it can be omitted, $-\ln\mathcal{L}$ is then given as
\begin{equation}
    -\ln\mathcal{L} = -\sum_{i=1}^{N}\ln(\frac{\omega(\xi_i)}{\int \omega(\xi) \epsilon(\xi) d\xi}).
\end{equation}

So the likelihood function is
\begin{equation}
\begin{aligned}
    -\ln\mathcal{L} 
    &= -\sum_{i=1}^{N}\ln(\frac{\omega(\xi_i)}{\int \omega(\xi) \epsilon(\xi) d\xi}) \\
    &= -\sum_{i=1}^{N} \ln (\frac{d\sigma}{d\Phi}/\sigma).
\end{aligned}
\end{equation}

The background contribution is estimated using events in the  $\Sigma$ sideband region and is subtracted from the log-likelihood function in the $\Sigma$ signal region, i.e.,
\begin{equation}
   -\ln\mathcal{L} = -(\ln\mathcal{L}_{\mathrm{data}} - \ln\mathcal{L}_{\mathrm{bkg}}).
\end{equation}

The observed cross-section $\sigma$ is calculated by the MC integral is given as 

\begin{equation}
\begin{aligned}
   \sigma_{\mathrm{obs}} &=\int \omega(\xi) \epsilon(\xi) d\xi \\
   &= \sum_{i}\Delta \xi_i \omega(\xi_i) \epsilon(\xi_i) \\
   &= \frac{1}{N_{\mathrm{gen}}}\sum_{i} N_{\mathrm{gen}}\Delta \xi_i \omega(\xi_i) \epsilon(\xi_i).
\end{aligned}
\end{equation}
Because $N_{\xi_i}\equiv N_{\mathrm{gen}}\Delta \xi_i \epsilon(\xi_i)$, the observed cross-section is then given as 
\begin{equation}
   \sigma_{\mathrm{obs}} = \frac{1}{N_{\mathrm{gen}}}\sum_{i}N_{\xi_i}\omega(\xi_i)
   = \frac{1}{N_{\mathrm{gen}}}\sum_{k=1}^{N_{\mathrm{acc}}}\omega(\xi_i),
\end{equation}
where $N_{\mathrm{gen}}$ is the total number of MC events, $N_{\mathrm{acc}}$ is the number of MC events satisfied event selections, and $k$ denotes the $k$-th event. 

The construction of the probability density function, the calculation of the fit fraction, and the corresponding statistical uncertainty for each component follow Ref.~\cite{Oset:2001cn}. The combined branching fraction is obtained according to
\begin{equation}
\begin{aligned}
    &\BR ( J/\psi \to \rm{X} \bar{\Sigma}^0\to \Lambda\bar{\Sigma}^0\eta )
 = \frac{N_{\rm X}}{N_{ \jpsi } \cdot \BR_{\rm sub}\cdot \epsilon_{\rm X}}\\
& = \frac{N_{\rm total}}{N_{ \jpsi } \cdot \BR_{\rm sub} \cdot \epsilon_{\rm total}}\cdot\frac{\epsilon_{\rm total}\cdot N_{\rm X}}{\epsilon_{\rm X}\cdot N_{\rm total}}
\end{aligned}
\end{equation}
Here, $\BR_{\rm sub}$ denotes the product branching fraction of $\BR ( \Sigma ^0 \to \gamma \Lambda )\cdot \BR ^2( \Lambda \to p \pi ^-) \cdot \BR ( \eta \to \gamma \gamma)$, $N_{J/\psi}$, $N_{\rm total}$ and $N_{\rm X}$ represent the number of $J/\psi$, PWA data and Resonance $\rm X$, $\epsilon_{\rm X}$ is the detection efficiency for decays involving the resonance $\rm X$, while $\epsilon_{\rm total}$ is the total detection efficiency. The equation simplifies to
\begin{equation}
    \BR ( J/\psi \to \mathrm{X}\bar{\Sigma}^0\to \Lambda\bar{\Sigma}^0\eta )
  = (J/\psi \to \Lambda\bar{\Sigma}^0\eta)\cdot FF_{\rm X},
\end{equation}
where $FF_{\rm X}$ is the fit fraction of resonance $\rm X$.

Based on TF-PWA, we perform a combined fit for the charge-conjugate modes $J/\psi \to \Lambda \bar{\Sigma}^0\eta $ and $J/\psi \to \bar{\Lambda}\Sigma^0\eta $. In the fit, the likelihood of each decay mode is calculated using the MC integral and data sample, while these two decay modes share the same set of resonances' parameters and waves' parameters, namely the amplitude and phase of each wave. The total likelihood value is then constructed by the product of the likelihood value for each mode.

\begin{table}[htbp]
    \centering
    \caption{The possible intermediate states of $ \Lambda ^*$ states~(PDG~\cite{ParticleDataGroup:2024cfk}).}
    \label{tab:eta_pwa_resonance}
    \begin{tabular}{ccccccccc}
        \toprule
        Resonance         & $M$(MeV$/c^2 $)               & $\Gamma$(MeV)          & $J^{P}$     & Existence     \\
        \hline
        $ \Lambda (1600)$       & $1600\pm30$       & $200\pm50$      & $1/2^+$        & ****           \\
        $ \Lambda (1670)$       & $1674\pm4$        & $30\pm5$        & $1/2^-$        & ****           \\
        $ \Lambda (1690)$       & $1690\pm5$        & $70\pm10$       & $3/2^-$        & ****             \\
        $ \Lambda (1710)$       & $1713\pm13$       & $180\pm40$      & $1/2^+$        & *               \\
        $ \Lambda (1800)$       & $1800\pm50$       & $200\pm50$      & $1/2^-$        & ***            \\
        $ \Lambda (1810)$       & $1790\pm50$       & $110\pm60$      & $1/2^+$        & ***                  \\
        $ \Lambda (1890)$       & $1890\pm20$       & $120\pm40$      & $3/2^+$        & ****            \\
        \toprule
    \end{tabular}
\end{table}

\begin{figure*}[htbp]
    \centering
    \begin{overpic}[width=0.32\textwidth]{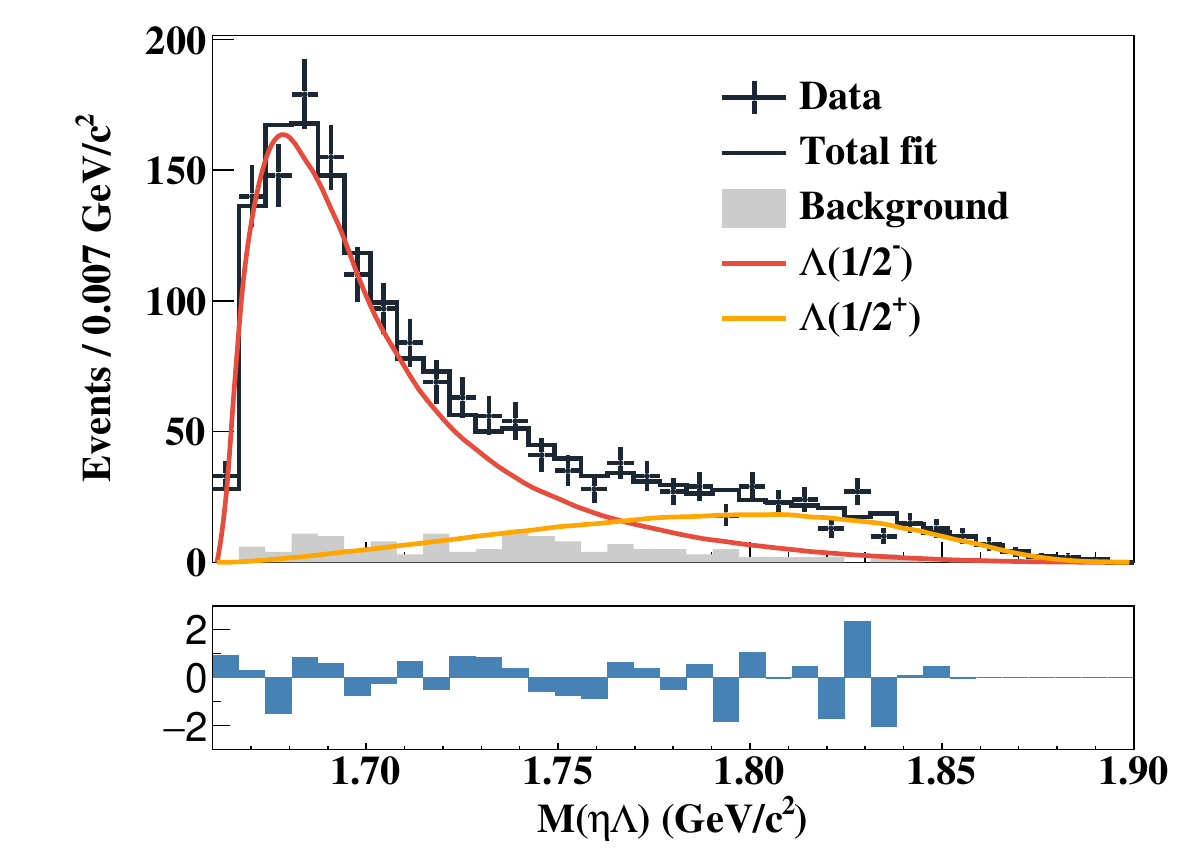}
        \put(30,62){$(a)$}
    \end{overpic}
    \begin{overpic}[width=0.32\textwidth]{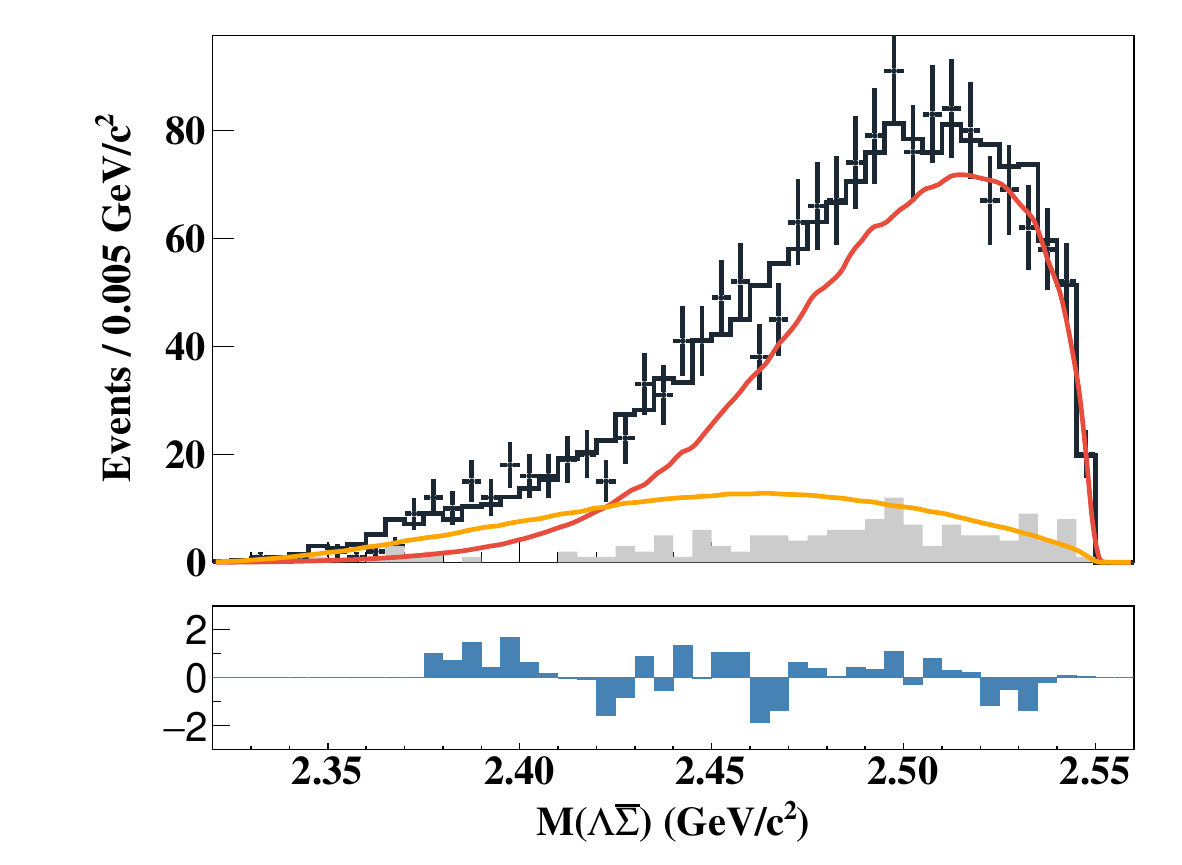}
        \put(85,62){$(b)$}
    \end{overpic}
    \begin{overpic}[width=0.32\textwidth]{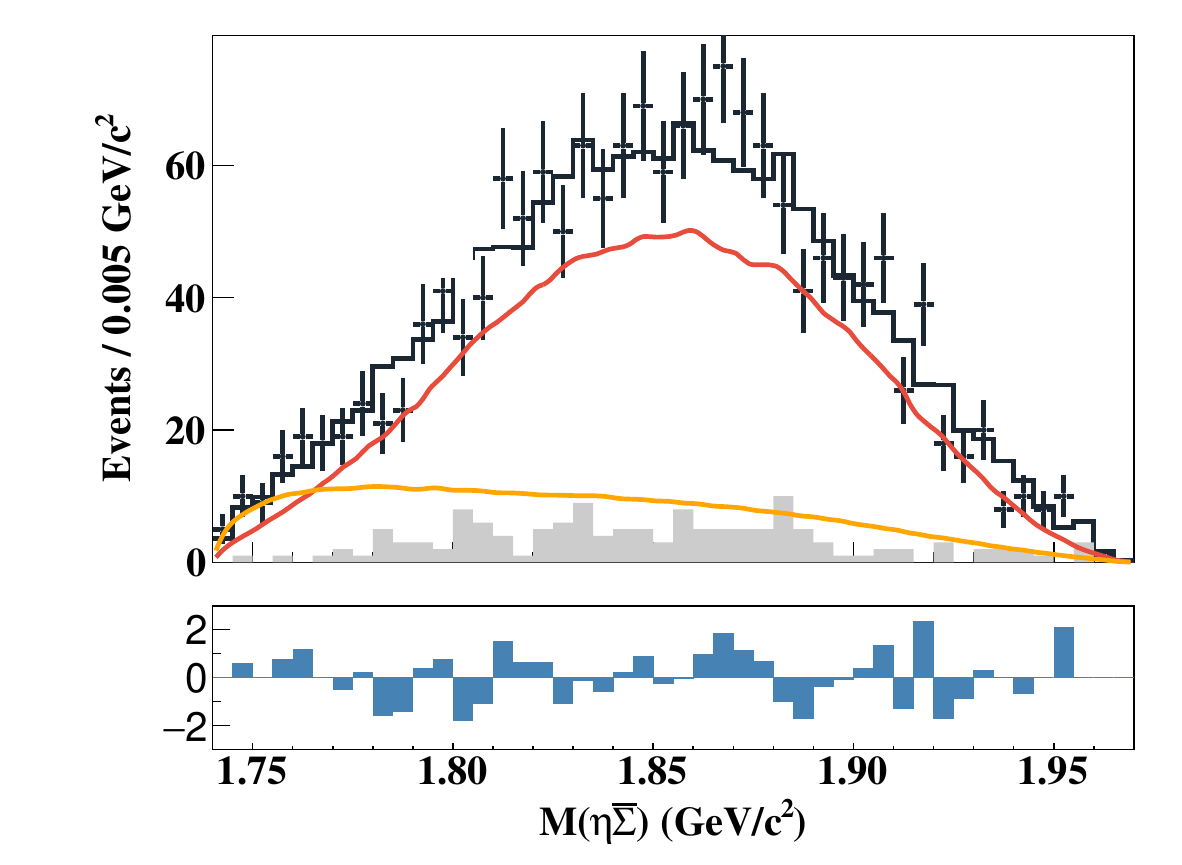}
        \put(82,62){$(c)$}
    \end{overpic}
        \caption{The distributions of (a) $ M ( \eta  \Lambda )$,  (b) $ M ( \Lambda \bar{\Sigma}^0)$ and (c)$M ( \eta \bar{\Sigma}^0  )$. The black lines are the total PWA results, the gray histograms represent the background, while the red lines are $\Lambda(1/2^-)$ and the orange lines are $\Lambda(1/2^+)$.}
        \label{fig:pwa_m}
\end{figure*}
\begin{figure*}[t]
    \centering
    \begin{overpic}[width=0.32\textwidth]{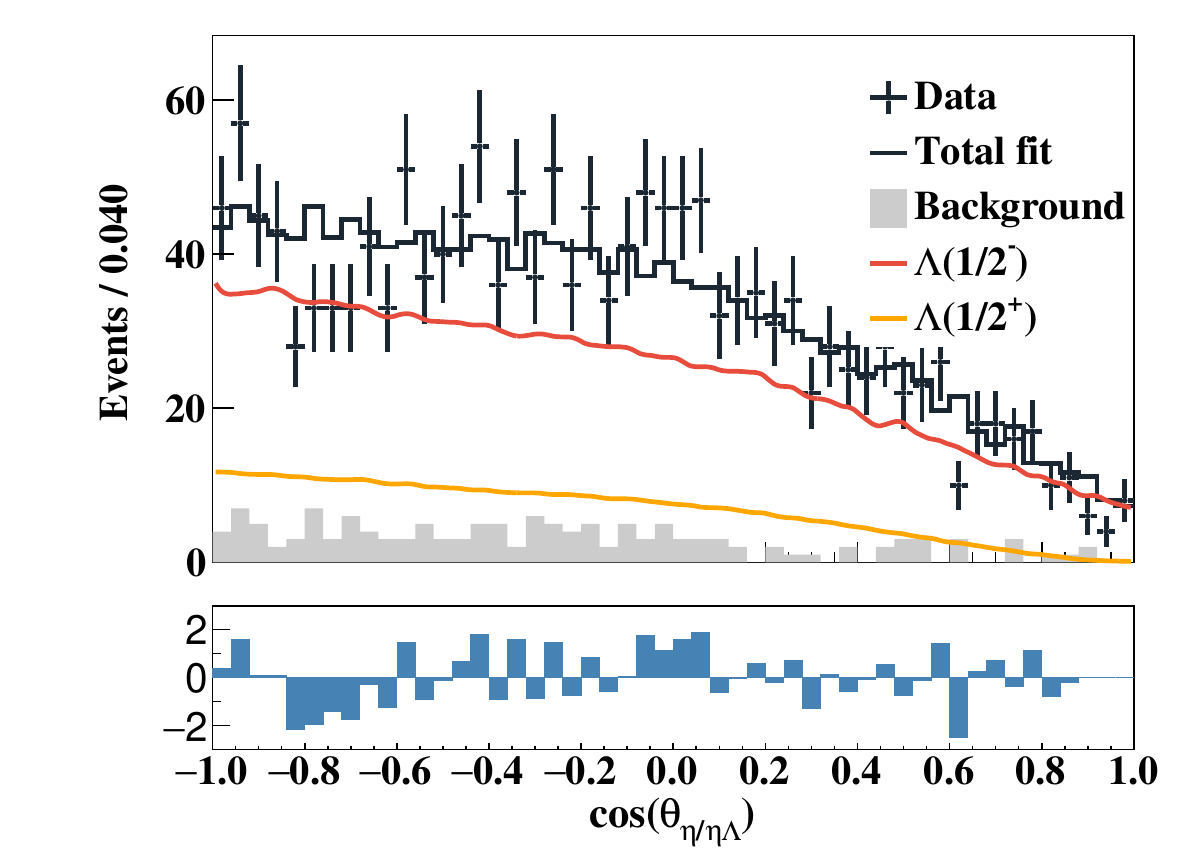}
        \put(26,62){$(a)$}
    \end{overpic}
    \begin{overpic}[width=0.32\textwidth]{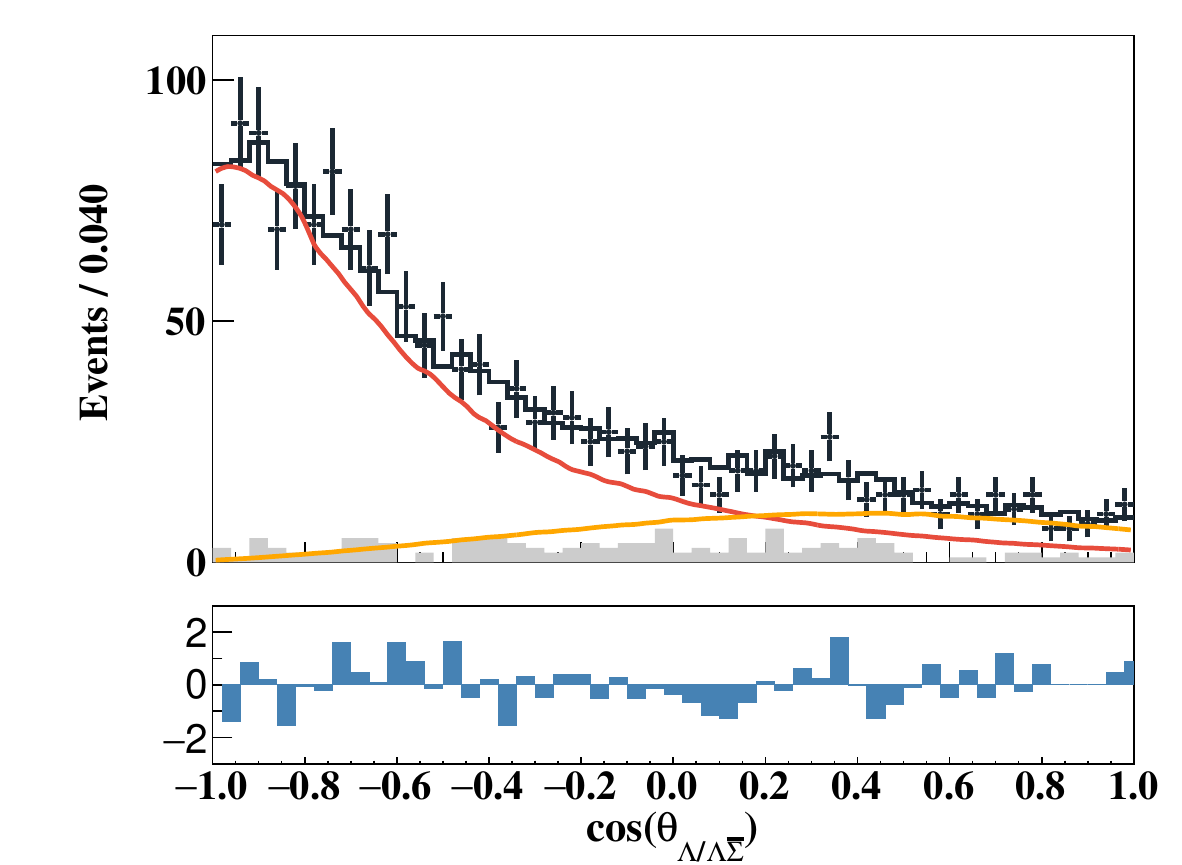}
        \put(78,62){$(b)$}
    \end{overpic}
    \begin{overpic}[width=0.32\textwidth]{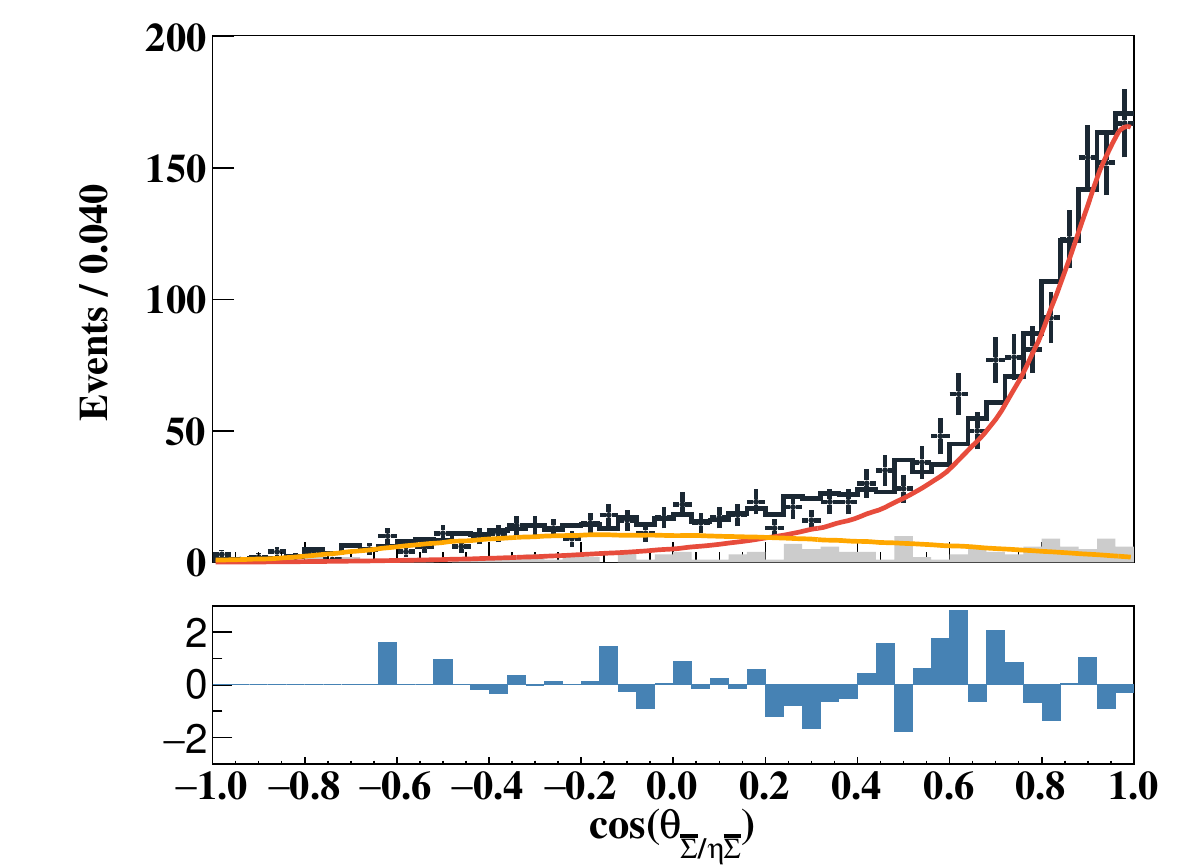}
        \put(78,62){$(c)$}
    \end{overpic}
    \caption{The distributions of (a) $ \cos{\theta_{\eta/\eta\Lambda}}$, (b) $ \cos{\theta_{\Lambda/\Lambda\bar{\Sigma}}}$ and (c)$ \cos{\theta_{\bar{\Sigma}/\eta\bar{\Sigma}}}$. The black lines are the total PWA results. The gray histograms represent the background. The red and orange lines show the components of $\Lambda(1/2^-)$ and $\Lambda(1/2^+)$, respectively.}
    \label{fig:pwa_cos}
\end{figure*}

\subsection{PWA results}

To obtain a good description of the data requires testing for the presence and significance of numerous resonances and their combinations in the PWA parameterization. The goal is to minimize the total number of required resonances, i.e., to find the simplest parameterization that describes the data within uncertainties. However, the number of possible resonance combinations
is usually far beyond what can be tested in the traditional way, such as manually inserting Breit-Wigner resonant terms into simple parameterizations.

Since no obvious structure on the $M(\Sigma^{0*} \eta)$ is evident as shown in Fig.~\ref{fig:pwa_m}(c), we only consider the excited Lambda hyperon ($\Lambda^* \to \Lambda \eta$) in the PWA.
The $\Lambda^*$ candidates allowed, given the constraints of the limited phase space, are listed in Table~\ref{tab:eta_pwa_resonance}.
Since many of them are not well established, extensive checks have been performed to verify the reliability of this analysis. 
In particular, we tested alternative spin-parity ($J^P$) assignments for each state to find the optimized solution.
Only components with a statistical significance exceeding $5\sigma$ are kept in the baseline solution.

It is found that two resonant contributions, with $J^P=1/2^-$ and $J^P=1/2^+$, are sufficient to 
describe the data well.
Comparisons between data and the PWA fit projections for the distributions of  $M(\eta\Lambda)$, $M(\eta\bar{\Sigma})$, and $M(\Lambda\bar{\Sigma})$ are shown in Fig.~\ref{fig:pwa_m}. 
The fit projections of the angular distributions are shown in Fig.~\ref{fig:pwa_cos}, demonstrating good consistency between data and the PWA results. 
The extracted parameters and statistical significances for each resonance are presented in Table~\ref{tab:pwa}. The statistical significance is evaluated using the change in the log-likelihood value (denoted as $\Delta S$ ) and the corresponding number of degrees of freedom (denoted as $\Delta $ndf) in the fits with and without the specific component. 
A resonance with spin-parity assignment of $1/2^-$ is observed near 1.67 $\mathrm{GeV}/c^2$ in the $\eta\Lambda$ mass spectrum. The measured mass and width are in agreement with the values of the $\Lambda(1670)$ in the PDG listings.
The second observed resonance could be identified as the $\Lambda(1810)$ based on its $1/2^+$ spin-parity. 
Its mass is about 90~$\mathrm{MeV}/c^2$ higher than the value of the $\Lambda(1810)$ in the PDG, 
but the difference is within two standard deviations. Its mass and width also match those of the $\Lambda(1890)$, 
which has $3/2^+$ spin-parity in the PDG.
To test this possibility, we assign $J^P=3/2^+$ to the second resonance and find that the fit quality is just slightly worse with a change of log-likelihood value of $\Delta S=5$. Thus, while we cannot definitively claim that the second resonant component is the $\Lambda(1810)$, the $\Lambda(1810)$ possibility is slightly favored over the $\Lambda(1890)$. The measured masses, widths, and product branching fractions for each component are summarized in Table~\ref{tab:pwa}.

\begin{table*}[t]
    \centering
    \caption{The resonant parameters, the fitted fractions and the detection efficiencies for each component in the baseline solution, where the first uncertainties are statistical and the second systematic.}
    \label{tab:pwa}
    \begin{tabular}{cccccccc}
        \toprule
        Resonance    &$M$(MeV/$c^2$)   &$\Gamma$(MeV)    &$\BR(J/\psi\rightarrow \Lambda^*\bar{\Sigma}^0\rightarrow\Lambda\bar{\Sigma}^0\eta)$  & Significance \\
        \hline
        $ \Lambda (1670)$    &$1668.8\pm3.1 \pm21.2$       & $52.7\pm4.2 \pm17.8$     &$(2.47\pm0.09\pm0.44 )\times 10^{-5}$    &$>20\sigma$      \\
        $ \Lambda (1810)$    &$1881.5\pm16.5 \pm20.3$      & $82.4\pm18.2 \pm8.9$     &$(1.02\pm0.13\pm0.14 )\times 10^{-5}$    &10.9$\sigma$       \\
        \hline\hline
    \end{tabular}
\end{table*}

\subsection{Branching fraction measurement of $J/\psi\rightarrow\Lambda\bar{\Sigma}^0\eta$}

\begin{figure}[htbp]
\centering
\subfigure{\includegraphics[scale=0.35]{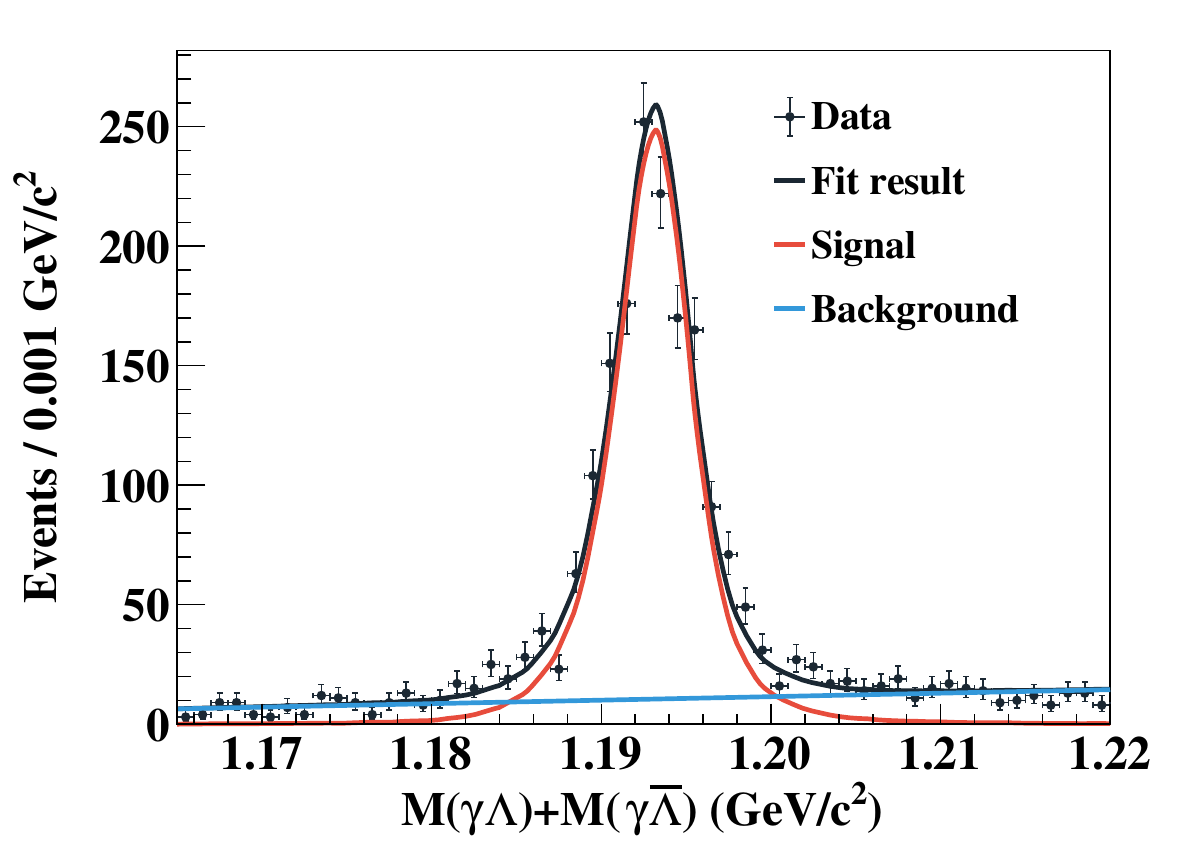}}
\caption{Fit to the invariant mass distribution of $\gamma\Lambda+\gamma\bar{\Lambda}$. Dots with error bars represent the data, the red lines are the MC signal shape, the blue lines are the background shape, and the black lines are the total fit result.}
\label{fit_br}
\end{figure}

To extract the number of $J/\psi\rightarrow\Lambda\bar{\Sigma}^0\eta$ events, an unbinned maximum likelihood fit is performed to the $\gamma\bar{\Lambda}$ invariant mass spectrum. 
The signal component is modeled with the MC simulated signal shape convolved with a Gaussian function to account for a
potential difference in the mass resolution between data and MC simulation. The combinatorial background is parameterized by a first-order Chebychev polynomial function. The resulting fit result, presented in Fig.~\ref{fit_br}, 
yields $N^{\rm{obs}}_{ \Sigma }=1553 \pm 44$ signal events.

The branching fraction is determined as:
\begin{equation}
\begin{aligned}
&\BR ( J/\psi \to \Lambda\bar{\Sigma}^0\eta )\\
& = \frac{N^{\rm{obs}}_{ \Sigma}}{N_{ \jpsi } \cdot \BR ( \Sigma ^0 \to \gamma \Lambda )\cdot \BR ^2( \Lambda \to p \pi ^-) \cdot \BR ( \eta \to \gamma \gamma ) \cdot \epsilon},
\end{aligned}
\end{equation}
where $N_{\Sigma}^{\rm{obs}}$ is the signal yield determined in the fit, $\varepsilon$ is the detection efficiency for the decay, $\mathcal{B}(\Sigma\to\gamma\Lambda), \mathcal{B}(\Lambda\to p\pi^-)$ and $\mathcal{B}(\eta\to \gamma\gamma)$ are the corresponding branching fractions from the PDG~\cite{ParticleDataGroup:2024cfk}, and $N_{J/\psi}$ is the total number of $J/\psi$ events in the data,
 $(10087\pm44)\times10^{6}$~\cite{BESIII_2021_njpsi}. With a detection efficiency of $2.78\%$, the branching fraction of $J/\psi \to \Lambda\bar{\Sigma}^0\eta$ is calculated to be $(3.44 \pm 0.11 ) \times 10^{-5}$.
Using the same approach to determine the detection efficiency, the product branching fraction for each component is given in Table~\ref{tab:tabC}.
\begin{table}[htbp]
   \centering
    \caption {The signal yields ($N^{\rm{obs}}$), detection efficiencies and branching fractions for each product branching fraction, where the uncertainties are statistical only. }
    \label{tab:tabC}
   \begin{tabular}{cccc}
        \toprule
     Resonance    &$N^{\mathrm{obs}} $  &$\epsilon(\%) $    &$\BR(\times10^{-5})$  \\
        \hline
        $\Lambda(1670) $     & $1257\pm45$          & 3.14     &$2.47\pm0.09$      \\
        $\Lambda(1810) $     & $353\pm45$           & 2.13     &$1.02\pm0.13$     \\
        \hline\hline
    \end{tabular}
    \end{table}

\section{Systematic uncertainties}

\subsection{Uncertainties from event selection}

Table~\ref{sys_br} lists all the systematic uncertainty sources
 and values. These systemtatic uncertainties are statistically independent and can be summed up in quadrature. The different sources of systematic  uncertainties for the measurement of the branching fraction are considered and described below.
 \begin{table}[hb]
    \centering
    \caption{Systematic uncertainties in the measurement of the branching fraction for $ \jpsi \to  \Lambda \bar{\Sigma}^0\eta$.}
    \label{sys_br}
    \begin{tabular}{l|c}
        \toprule
        Source                                                               & Uncertainty$ (\%)$ \\
        \hline
        $ \Lambda/\bar{\Lambda} $ reconstruction                         & 0.8            \\
        Photon detection                                           & 1.5            \\
        Kinematic fit                                                         & 1.7            \\
        Fit range                                                         & 1.7            \\
        Background shape                                                & 2.0            \\
        Quoted branching fractions                                          & 1.6            \\
        Number of $ \jpsi $ events                                            & 0.4            \\
        \hline
        Total                                                                 & 3.9            \\
        \hline\hline
    \end{tabular}
\end{table}

\begin{itemize}

\item{$\it \Lambda/\bar{\Lambda}$ reconstruction}: The efficiency of $ \Lambda / \bar{ \Lambda } $ reconstruction, incorporating both the MDC tracking and the $\Lambda(\bar{\Lambda})$ mass window requirement, is studied using a control sample of
$ \jpsi \to \Lambda \bar\Lambda  $ decays, and a  correction factor of $0.997\pm0.004$~\cite{BESIII:2017kqw} is applied to the efficiencies obtained from MC simulation. After efficiency correction we take 0.8$\%$ as the systematic uncertainty due to
 the $\Lambda/ \bar{ \Lambda } $ reconstruction.

\item{$\it$Photon detection}: The systematic uncertainty from the reconstruction of photons has been studied extensively in the process $e^+e^-\rightarrow\gamma\mu^+\mu^-$. 
The relative difference in efficiency between data and MC simulation, $0.5\%$, is assigned as the systematic uncertainty. Since there are three photons in the final state of interest, the uncertainty from photon detection is $1.5\%$.

\item{$\it$Kinematic fit}: To investigate the systematic uncertainty associated with the kinematic fit, the track helix parameter correction method~\cite{BESIII_2013_kmfitfit} is used. The difference in the detection efficiencies with and without  the helix correction is taken as the systematic uncertainty.

\item{$\it$Fit range}: The systematic uncertainty from the fit  range is estimated by changing the fit range by $\pm 3$ $\mathrm{MeV}/c^2$. The difference in the branching fraction with different fit ranges is taken as the systematic uncertainty.

\item{$\it$Background shape}: To estimate the systematic uncertainty due to choice of background shape, we change the order of the Chebychev polynomial from first order to second order in the fit. The maximum difference in the fitted signal yield, $1.9\%$,  is taken as the associated systematic uncertainty.

\item{$\it$Quoted branching fractions}: The uncertainties from the quoted branching fractions of the intermediate decays of $\Lambda \to p \pi^-$ and $\eta\to\gamma\gamma$ are quoted from the PDG~\cite{ParticleDataGroup:2024cfk}. 

\item{$\it$Number of $J/\psi$ events}: The total number of $J/\psi$ events is determined to be $(10087\pm44)\times10^6$ by counting inclusive hadronic events, and its uncertainty is 0.4\%~\cite{BESIII_2021_njpsi}.

\end{itemize}

\begin{table*}[ht]
    \centering
    \caption{The systematic uncertainties of the masses, widths and fitted fractions of $\Lambda(1670)$ and $\Lambda(1810)$.} 
    \label{tab:sys_pwa}
    \begin{tabular}{c|ccc|ccc}
        \toprule
        \multirow{2}{*}{Source} & \multicolumn{3}{c|}{$\Lambda(1670)$} & \multicolumn{3}{c}{$\Lambda(1810)$} \\
        & \textbf{$\Delta M(\mathrm{MeV}/c^2)$} & \textbf{$\Delta \Gamma(\mathrm{MeV})$} & \textbf{$\Delta \BR(\%)$} & \textbf{$\Delta M(\mathrm{MeV}/c^2)$} & \textbf{$\Delta \Gamma(\mathrm{MeV})$} & \textbf{$\Delta \BR(\%)$} \\  
        \hline
        Extra resonances & 3.9 &14.2  &17.2  &18.9  &7.2  &11.9 \\ 
        Background estimation    & 0.2 &0.5   &0.1  &0.9   &0.1  &1.3 \\ 
        Mass resolution  & 1.0 &3.1   &1.5  &2.8   &4.5  &5.3 \\ 
        Resonance parameterization  & 20.8 &10.3 &1.1  &6.9   &2.7  &4.0 \\ 
        \hline
        Total &21.2 &17.8 &17.3 &20.3 &8.9 &13.7 \\ 
        \hline\hline
    \end{tabular}
\end{table*}

\subsection{Uncertainties from the PWA}

The systematic uncertainties in the PWA, listed in Table~\ref{tab:sys_pwa}, are discussed in detail below. Assuming all sources are independent, the total systematic uncertainty is determined by adding them in quadrature.

\begin{itemize}
\item{$\it$Extra resonances}: To evaluate the effect on the PWA results from other possible components, the PWA is re-performed by adding extra resonances $\Lambda(\frac{3}{2}^-)$. The largest changes of the masses, widths, and fitted fractions of each resonance are taken as the systematic uncertainties.

\item{$\it$Background estimation}: The uncertainty due to the background estimation is evaluated by using different sideband regions, and changing the background level through changing the sideband normalization factors by one standard deviation. The sideband normalization factors are increased or decreased by one standard deviation and the maximum difference is taken as the systematic uncertainty.

\item{$\it$Mass resolution}: Uncertainties from the mass resolution are estimated by adding resolution in the PWA.
The impacts on the masses and widths of $ \Lambda (\frac{1}{2}^- )$, $ \Lambda (\frac{1}{2}^+ )$, and the branching fractions  of intermediate states are taken as the uncertainties from mass resolution.

\item{$\it$Resonance parameterization}: The uncertainties due to  the parameterizations of resonance states are estimated by replacing the constant width of the Breit-Wigner with the mass-dependent width $\Gamma(m)$ to the Breit-Wigner parameterization~\cite{ParticleDataGroup:2024cfk}.

\end{itemize}

\section{SUMMARY}

Based on a sample of $(10087\pm44)\times 10^6$ $J/\psi$ events collected with the BESIII detector, a PWA of $J/\psi \to \Lambda\bar{\Sigma}^0\eta$ is performed for the first time to investigate excited $\Lambda$ states. 
The data are well described by a fit with two resonances, the $\Lambda(1670)$ with $J^P=1/2^-$ and the $\Lambda(1810)$ with $J^P=1/2^+$.
The branching fraction of $J/\psi\to\Lambda\bar{\Sigma}^0\eta$ is also measured to be $\mathcal{B}(J/\psi\to\Lambda\bar{\Sigma}^0\eta) = (3.44 \pm 0.11 \pm 0.13$)$ \times 10^{-5}$.

\begin{acknowledgments}

The BESIII Collaboration thanks the staff of BEPCII (https://cstr.cn/31109.02.BEPC) and the IHEP computing center for their strong support. This work is supported in part by National Key R\&D Program of China under Contracts Nos. 2025YFA1613900, 2023YFA1606000, 2023YFA1606704; National Natural Science Foundation of China (NSFC) under Contracts Nos. 12225509, 12465015, 12575087, 11635010, 11935015, 11935016, 11935018, 12025502, 12035009, 12035013, 12061131003, 12192260, 12192261, 12192262, 12192263, 12192264, 12192265, 12221005, 12235017, 12342502, 12361141819; the Chinese Academy of Sciences (CAS) Large-Scale Scientific Facility Program; the Strategic Priority Research Program of Chinese Academy of Sciences under Contract No. XDA0480600; CAS under Contract No. YSBR-101; 100 Talents Program of CAS; The Institute of Nuclear and Particle Physics (INPAC) and Shanghai Key Laboratory for Particle Physics and Cosmology; ERC under Contract No. 758462; German Research Foundation DFG under Contract No. FOR5327; Istituto Nazionale di Fisica Nucleare, Italy; Knut and Alice Wallenberg Foundation under Contracts Nos. 2021.0174, 2021.0299, 2023.0315; Ministry of Development of Turkey under Contract No. DPT2006K-120470; National Research Foundation of Korea under Contract No. NRF-2022R1A2C1092335; National Science and Technology fund of Mongolia; Polish National Science Centre under Contract No. 2024/53/B/ST2/00975; STFC (United Kingdom); Swedish Research Council under Contract No. 2019.04595; U. S. Department of Energy under Contract No. DE-FG02-05ER41374

\end{acknowledgments}

\bibliography{apssamp}

\end{document}

%% file: authorlist_2025-09-11.tex
M.~Ablikim$^{1}$\BESIIIorcid{0000-0002-3935-619X},
M.~N.~Achasov$^{4,b}$\BESIIIorcid{0000-0002-9400-8622},
P.~Adlarson$^{81}$\BESIIIorcid{0000-0001-6280-3851},
X.~C.~Ai$^{86}$\BESIIIorcid{0000-0003-3856-2415},
C.~S.~Akondi$^{31A,31B}$\BESIIIorcid{0000-0001-6303-5217},
R.~Aliberti$^{39}$\BESIIIorcid{0000-0003-3500-4012},
A.~Amoroso$^{80A,80C}$\BESIIIorcid{0000-0002-3095-8610},
Q.~An$^{77,64,\dagger}$,
Y.~H.~An$^{86}$\BESIIIorcid{0009-0008-3419-0849},
Y.~Bai$^{62}$\BESIIIorcid{0000-0001-6593-5665}
% ...
\\
\vspace{0.2cm}
(BESIII Collaboration)\\
\vspace{0.2cm} {\it
$^{1}$ Institute of High Energy Physics, Beijing 100049, People's Republic of China\\
% ...
$^{4}$ Budker Institute of Nuclear Physics SB RAS (BINP), Novosibirsk 630090, Russia\\
% ...
$^{31}$ INFN Sezione di Ferrara, (A)INFN Sezione di Ferrara, I-44122, Ferrara, Italy; (B)University of Ferrara, I-44122, Ferrara, Italy\\
% ...
$^{39}$ Johannes Gutenberg University of Mainz, Johann-Joachim-Becher-Weg 45, D-55099 Mainz, Germany\\
% ...
$^{62}$ Southeast University, Nanjing 211100, People's Republic of China\\
% ...
$^{64}$ State Key Laboratory of Particle Detection and Electronics, Beijing 100049, Hefei 230026, People's Republic of China\\
% ...
$^{77}$ University of Science and Technology of China, Hefei 230026, People's Republic of China\\
% ...
$^{80}$ University of Turin and INFN, (A)University of Turin, I-10125, Turin, Italy; (B)University of Eastern Piedmont, I-15121, Alessandria, Italy; (C)INFN, I-10125, Turin, Italy\\
$^{81}$ Uppsala University, Box 516, SE-75120 Uppsala, Sweden\\
% ...
$^{86}$ Zhengzhou University, Zhengzhou 450001, People's Republic of China\\

\vspace{0.2cm}
$^{\dagger}$ Deceased\\
% ...
$^{b}$ Also at the Novosibirsk State University, Novosibirsk, 630090, Russia\\
% ...

}

%% file: apssamp.bib
@article{Sarantsev:2019xxm,
    author = "Sarantsev, A. V. and Matveev, M. and Nikonov, V. A. and Anisovich, A. V. and Thoma, U. and Klempt, E.",
    title = "{Hyperon II: Properties of excited hyperons}",
    doi = "10.1140/epja/i2019-12880-5",
    journal = "Eur. Phys. J. A",
    volume = "55",
    number = "10",
    pages = "180",
    year = "2019"
}

@article{Kamano:2015hxa,
    author = "Kamano, H. and Nakamura, S. X. and Lee, T. -S. H. and Sato, T.",
    title = "{Dynamical coupled-channels model of $K^- p$ reactions. II. Extraction of $\Lambda^*$ and $\Sigma^*$ hyperon resonances}",
    doi = "10.1103/PhysRevC.92.025205",
    journal = "Phys. Rev. C",
    volume = "92",
    number = "2",
    pages = "025205",
    year = "2015"
}

@article{Koniuk:1979vy,
    author = "Koniuk, Roman and Isgur, Nathan",
    title = "{Baryon Decays in a Quark Model with Chromodynamics}",
    reportNumber = "Print-79-1030 (TORONTO)",
    doi = "10.1103/PhysRevD.21.1868",
    journal = "Phys. Rev. D",
    volume = "21",
    pages = "1868",
    year = "1980"
}

@article{Oset:2001cn,
    author = "Oset, E. and Ramos, A. and Bennhold, C.",
    title = "{Low lying S = -1 excited baryons and chiral symmetry}",
    doi = "10.1016/S0370-2693(01)01523-4",
    journal = "Phys. Lett. B",
    volume = "527",
    pages = "99--105",
    year = "2002",
}

@article{Zou:2013af,
    author = "Zou, B. S.",
    editor = "Juli\'a-D\'\i{}az, Bruno and Magas, Volodymyr and Oset, Eulogio and Parre\~no, Assumpta and Polls, Artur and Tol\'os, Laura and Vida\~na, Isaac and Ramos, \`Angels",
    title = "{Hadron spectroscopy from strangeness to charm and beauty}",
    doi = "10.1016/j.nuclphysa.2013.01.001",
    journal = "Nucl. Phys. A",
    volume = "914",
    pages = "454--460",
    year = "2013"
}

@article{Zhang:2013sva,
    author = "Zhang, H. and Tulpan, J. and Shrestha, M. and Manley, D. M.",
    title = "{Multichannel parametrization of $\bar K N$ scattering amplitudes and extraction of resonance parameters}",
    doi = "10.1103/PhysRevC.88.035205",
    journal = "Phys. Rev. C",
    volume = "88",
    number = "3",
    pages = "035205",
    year = "2013"
}

@article{Belle:2020xku,
    author = "Lee, J. Y. and others",
    collaboration = "Belle",
    title = "{Measurement of branching fractions of  $\Lambda_{c}^{+} \rightarrow \eta\Lambda\pi^{+}$, $\eta \Sigma^{0} \pi^{+}$, $\Lambda(1670) \pi^{+}$, and $\eta \Sigma(1385)^{+}$}",
    doi = "10.1103/PhysRevD.103.052005",
    journal = "Phys. Rev. D",
    volume = "103",
    number = "5",
    pages = "052005",
    year = "2021"
}

@article{ParticleDataGroup:2024cfk,
    author = "Navas, S. and others",
    collaboration = "Particle Data Group",
    title = "{Review of particle physics}",
    doi = "10.1103/PhysRevD.110.030001",
    journal = "Phys. Rev. D",
    volume = "110",
    number = "3",
    pages = "030001",
    year = "2024"
}

@article{BESIII:2012xdg,
    author = "Ablikim, M. and others",
    collaboration = "BESIII",
    title = "{First observation of the isospin violating decay $J/\psi\rightarrow \Lambda\bar{\Sigma}^{0}+c.c.$}",
    doi = "10.1103/PhysRevD.86.032008",
    journal = "Phys. Rev. D",
    volume = "86",
    pages = "032008",
    year = "2012"
}

@article{DM2:1988bfq,
    author = "Jousset, J. and others",
    collaboration = "DM2",
    title = "{The $J/\psi \to$ Vector + Pseudoscalar Decays and the $\eta$, $\eta^\prime$ Quark Content}",
    reportNumber = "LAL-88-25",
    doi = "10.1103/PhysRevD.41.1389",
    journal = "Phys. Rev. D",
    volume = "41",
    pages = "1389",
    year = "1990"
}

@article{BESIII:2016kpv,
    author = "Ablikim, Medina and others",
    collaboration = "BESIII",
    title = "{Determination of the number of $J/\psi$ events with inclusive $J/\psi$ decays}",
    doi = "10.1088/1674-1137/41/1/013001",
    journal = "Chin. Phys. C",
    volume = "41",
    number = "1",
    pages = "013001",
    year = "2017"
}

@article{BESIII_2009_detector,
    author = "Ablikim, M. and others",
    collaboration = "BESIII",
    title = "{Design and Construction of the BESIII Detector}",
    doi = "10.1016/j.nima.2009.12.050",
    journal = "Nucl. Instrum. Meth. A",
    volume = "614",
    pages = "345--399",
    year = "2010"
}

@inproceedings{Yu:IPAC2016-TUYA01,
  author    = {C.~H.~Yu and others},
  title     = {BEPCII Performance and Beam Dynamics Studies on Luminosity},
  booktitle = {Proceedings of IPAC2016},
  year      = {2016},
  address   = {Busan, Korea},
  publisher  = {JACoW},
  doi       = {10.18429/JACoW-IPAC2016-TUYA01}
}

@article{BESIII:2020nme,
    author = "Ablikim, M. and others",
    collaboration = "BESIII",
    title = "{Future Physics Programme of BESIII}",
    reportNumber = "HEP-Physics-Report-BESIII-2019-12-13",
    doi = "10.1088/1674-1137/44/4/040001",
    journal = "Chin. Phys. C",
    volume = "44",
    number = "4",
    pages = "040001",
    year = "2020"
}

@article{BESIII_2019_detector,
    author = "Cao, P. and others",
    title = "{Design and construction of the new BESIII endcap Time-of-Flight system with MRPC Technology}",
    doi = "10.1016/j.nima.2019.163053",
    journal = "Nucl. Instrum. Meth. A",
    volume = "953",
    pages = "163053",
    year = "2020"
}

@article{GEANT4:2002zbu,
    author = "Agostinelli, S. and others",
    collaboration = "GEANT4",
    title = "{GEANT4--a simulation toolkit}",
    reportNumber = "SLAC-PUB-9350, FERMILAB-PUB-03-339, CERN-IT-2002-003",
    doi = "10.1016/S0168-9002(03)01368-8",
    journal = "Nucl. Instrum. Meth. A",
    volume = "506",
    pages = "250--303",
    year = "2003"
}

@article{Jadach:2000ir,
    author = "Jadach, S. and Ward, B. F. L. and Was, Z.",
    title = "{Coherent exclusive exponentiation for precision Monte Carlo calculations}",
    reportNumber = "CERN-TH-2000-087, UTHEP-99-09-01",
    doi = "10.1103/PhysRevD.63.113009",
    journal = "Phys. Rev. D",
    volume = "63",
    pages = "113009",
    year = "2001"
}

@article{Jadach:1999vf,
    author = "Jadach, S. and Ward, B. F. L. and Was, Z.",
    title = "{The Precision Monte Carlo event generator K K for two fermion final states in e+ e- collisions}",
    reportNumber = "DESY-99-106, CERN-TH-99-235, UTHEP-99-08-01",
    doi = "10.1016/S0010-4655(00)00048-5",
    journal = "Comput. Phys. Commun.",
    volume = "130",
    pages = "260--325",
    year = "2000"
}

@article{Lange:2001uf,
    author = "Lange, D. J.",
    editor = "Erhan, S. and Schlein, P. and Rozen, Y.",
    title = "{The EvtGen particle decay simulation package}",
    doi = "10.1016/S0168-9002(01)00089-4",
    journal = "Nucl. Instrum. Meth. A",
    volume = "462",
    pages = "152--155",
    year = "2001"
}

@article{Ping:2008zz,
    author = "Ping, R. G.",
    title = "{Event generators at BESIII}",
    doi = "10.1088/1674-1137/32/8/001",
    journal = "Chin. Phys. C",
    volume = "32",
    pages = "599",
    year = "2008"
}

@article{Chen:2000tv,
    author = "Chen, J. C. and Huang, G. S. and Qi, X. R. and Zhang, D. H. and Zhu, Y. S.",
    title = "{Event generator for J / psi and psi (2S) decay}",
    doi = "10.1103/PhysRevD.62.034003",
    journal = "Phys. Rev. D",
    volume = "62",
    pages = "034003",
    year = "2000"
}

@article{Yang:2014vra,
    author = "Yang, R. L. and Ping, R. G. and Chen, Hong",
    title = "{Tuning and Validation of the Lundcharm Model with $J/\psi$ Decays}",
    doi = "10.1088/0256-307X/31/6/061301",
    journal = "Chin. Phys. Lett.",
    volume = "31",
    pages = "061301",
    year = "2014"
}

@article{Richter-Was:1992hxq,
    author = "Richter-Was, E.",
    title = "{QED bremsstrahlung in semileptonic B and leptonic tau decays}",
    reportNumber = "CERN-TH-6746-92",
    doi = "10.1016/0370-2693(93)90062-M",
    journal = "Phys. Lett. B",
    volume = "303",
    pages = "163--169",
    year = "1993"
}

@article{Xu:2009zzg,
    author = "Xu, Min and others",
    title = "{Decay vertex reconstruction and 3-dimensional lifetime determination at BESIII}",
    doi = "10.1088/1674-1137/33/6/005",
    journal = "Chin. Phys. C",
    volume = "33",
    pages = "428--435",
    year = "2009"
}

@article{BESIII:2021cvv,
    author = "Ablikim, Medina and others",
    collaboration = "BESIII",
    title = "{Measurement of $\Lambda$ baryon polarization in $e^+e^-\rightarrow\Lambda\bar\Lambda$ at $\sqrt{s} = 3.773$ GeV}",
    doi = "10.1103/PhysRevD.105.L011101",
    journal = "Phys. Rev. D",
    volume = "105",
    number = "1",
    pages = "L011101",
    year = "2022"
}

@article{BESIII_2021_njpsi,
    author = "Ablikim, M. and others",
    collaboration = "BESIII",
    title = "{Number of $J/\psi$ events at BESIII}",
    doi = "10.1088/1674-1137/ac5c2e",
    journal = "Chin. Phys. C",
    volume = "46",
    number = "7",
    pages = "074001",
    year = "2022"
}

@article{BESIII:2017kqw,
    author = "Ablikim, Medina and others",
    collaboration = "BESIII",
    title = "{Study of $J/\psi$ and $\psi(3686)$ decay to $\Lambda\bar{\Lambda}$ and $\Sigma^0\bar{\Sigma}^0$ final states}",
    doi = "10.1103/PhysRevD.95.052003",
    journal = "Phys. Rev. D",
    volume = "95",
    number = "5",
    pages = "052003",
    year = "2017"
}

@article{BESIII:2012pbg,
    author = "Ablikim, M. and others",
    collaboration = "BESIII",
    title = "{Determination of the number of $J/\psi$ events with $J/\psi \rightarrow \, inclusive$ decays}",
    reportNumber = "2012-0173",
    doi = "10.1088/1674-1137/36/10/001",
    journal = "Chin. Phys. C",
    volume = "36",
    pages = "915--925",
    year = "2012"
}

@article{Wang:2020giv,
    author = "Wang, Mengzhen and Jiang, Yi and Liu, Yinrui and Qian, Wenbin and Lyu, Xiaorui and Zhang, Liming",
    title = "{A novel method to test particle ordering and final state alignment in helicity formalism}",
    doi = "10.1088/1674-1137/abf139",
    journal = "Chin. Phys. C",
    volume = "45",
    number = "6",
    pages = "063103",
    year = "2021"
}

@article{Chung:1997jn,
    author = "Chung, S. U.",
    title = "{A General formulation of covariant helicity coupling amplitudes}",
    reportNumber = "BNL-QGS-97-051",
    doi = "10.1103/PhysRevD.57.431",
    journal = "Phys. Rev. D",
    volume = "57",
    pages = "431--442",
    year = "1998"
}

@article{Chung:1993da,
    author = "Chung, S. U.",
    title = "{Helicity coupling amplitudes in tensor formalism}",
    doi = "10.1103/PhysRevD.56.4419",
    journal = "Phys. Rev. D",
    volume = "48",
    pages = "1225--1239",
    year = "1993"
}

@article{Chung:1995dx,
    author = "Chung, S. U. and Brose, J. and Hackmann, R. and Klempt, E. and Spanier, S. and Strassburger, C.",
    title = "{Partial wave analysis in K matrix formalism}",
    doi = "10.1002/andp.19955070504",
    journal = "Annalen Phys.",
    volume = "4",
    pages = "404--430",
    year = "1995"
}

@article{BESIII:2019dme,
    author = "Ablikim, M. and others",
    collaboration = "BESIII",
    title = "{Partial wave analysis of $\psi(3686)\to K^{+}K^{-}\eta$}",
    doi = "10.1103/PhysRevD.101.032008",
    journal = "Phys. Rev. D",
    volume = "101",
    number = "3",
    pages = "032008",
    year = "2020"
}

@article{BESIII_2013_kmfitfit,
    author = "Ablikim, M. and others",
    collaboration = "BESIII",
    title = "{Search for hadronic transition $χ_{cJ} → η_cπ^+π^-$ and observation of $χ_{cJ} → K\overline{K}πππ$}",
    doi = "10.1103/PhysRevD.87.012002",
    journal = "Phys. Rev. D",
    volume = "87",
    number = "1",
    pages = "012002",
    year = "2013"
}

@article{Jiang:2024vbw,
    author = "Jiang, Yi",
    title = "{Amplitude analysis tools at BESIII}",
    doi = "10.1393/ncc/i2024-24203-0",
    journal = "Nuovo Cim. C",
    volume = "47",
    number = "4",
    pages = "203",
    year = "2024"
}
